\documentclass[aps,prd,onecolumn,groupedaddress,showpacs,nofootinbib,amssymb]{revtex4}
\usepackage[dvips]{graphicx}
\usepackage{amssymb}
\usepackage{amsmath}
\usepackage{graphicx,,color}
\usepackage{amsfonts}
\usepackage{bm}
\usepackage{cancel}
\usepackage{comment}
\newcommand\be{\begin{equation}}
\newcommand\ee{\end{equation}}

\allowdisplaybreaks[4]

\begin{document}

\tolerance=5000

\title{Causal Limit of Neutron Star Maximum Mass in $f(R)$
Gravity in View of GW190814}
\author{A.V. Astashenok$^{1}$, S. Capozziello$^{2,3,4,5}$\thanks{capozzie@na.infn.it}
S.D.~Odintsov,$^{6,7,8}$\,\thanks{odintsov@ieec.uab.es}
V.K.~Oikonomou,$^{5,9}$\,\thanks{v.k.oikonomou1979@gmail.com}}
\affiliation{$^{1)}$ Institute of Physics, Mathematics and IT, I.
Kant Baltic Federal University, Kaliningrad, 236041 Russia,\\
$^{2)}$ Dipartimento di Fisica, ''E. Pancini'' Universit`a
''Federico II'' di Napoli, Compl. Univ. Monte S. Angelo
Ed. G, Via Cinthia, I-80126 Napoli, Italy,\\
$^{3)}$INFN Sez. di Napoli, Compl. Univ. Monte S. Angelo Ed. G,
Via Cinthia, I-80126 Napoli, Italy,
\\
$^{4)}$ Scuola Superiore Meridionale, Largo S. Marcellino 10, I-80138 Napoli, Italy,\\
$^{5)}$ Laboratory for Theoretical Cosmology, Tomsk State
University of Control Systems and Radioelectronics (TUSUR), 634050 Tomsk,
Russia,\\
$^{6)}$ ICREA, Passeig Luis Companys, 23, 08010 Barcelona, Spain\\
$^{7)}$ Institute of Space Sciences (IEEC-CSIC) C. Can Magrans
s/n,
08193 Barcelona, Spain\\
 $^{8)}$ Tomsk State Pedagogical University, 634061, Tomsk, Russia \\
 $^{9)}$ Department of Physics,
Aristotle University of Thessaloniki, Thessaloniki 54124, Greece.
}

\date{\today}
\tolerance=5000

\begin{abstract}
We investigate the causal limit of maximum mass for stars in the
framework of $f(R)$ gravity. We choose a causal equation of state,
with variable speed of sound, and with the transition density and
pressure corresponding to the SLy equation of state. The
transition density is chosen to be equal to twice the saturation
density $\rho_t\sim 2\rho_0$, and also  the analysis is performed
for the transition density, chosen to be equal to the saturation
density $\rho_t\sim \rho_0$. We examine numerically the combined
effect of the stiff causal equation of state and of the sound
speed  on the maximum mass of static neutron stars, in the context
of Jordan frame of $f(R)$ gravity. This  yields the most extreme
upper bound for neutron star masses in the context of extended
gravity. As we will evince for the case of $R^2$ model,  the upper
causal mass limit lies within, but not deeply in, the mass-gap
region, and is marginally the same with the general relativistic
causal maximum mass, indicating that the $\sim 3M_{\odot}$ general
relativistic limit is respected. In view of the modified gravity
perspective for the secondary component of the GW190814 event, we
also discuss the strange star possibility. Using several well
established facts for neutron star physics and the Occam's razor
approach, although the strange star is exciting, for the moment,
it remains a possibility for describing the secondary component of
GW190814. We underpin the fact that the secondary component of the
compact binary GW190814 is probably a neutron star, a black hole
or even a rapidly rotating neutron star, but not a strange star.
We also discuss, in general, the potential role of the extended
gravity description for the binary merging.
\end{abstract}

%PACS numbers: 04.50.Kd, 95.36.+x, 98.80.-k, 98.80.Cq
\pacs{04.50.Kd, 95.36.+x, 98.80.-k, 98.80.Cq,11.25.-w}

\maketitle

\section{Introduction}

Neutron stars (NSs) are currently in the epicenter of scientific
interest, since they are truly laboratories in the sky, for
many scientific purposes, like nuclear physics
\cite{Tolos:2020aln,Lattimer:2012nd,Steiner:2011ft,Horowitz:2005zb,Watanabe:2000rj,Shen:1998gq,Xu:2009vi,Hebeler:2013nza,Mendoza-Temis:2014mja,Ho:2014pta,Kanakis-Pegios:2020kzp},
particle physics
\cite{Buschmann:2019pfp,Safdi:2018oeu,Hook:2018iia,Edwards:2020afl}
and theoretical astrophysics
\cite{Sedrakian:2015krq,Khadkikar:2021yrj,Sedrakian:2006zza,Sedrakian:2018kdm,Bauswein:2020kor,Vretinaris:2019spn,Bauswein:2020aag,Bauswein:2017vtn}.
NSs have been thoroughly studied in the last 40 years, and the
latest LIGO-Virgo astronomical observations of gravitational waves
emitted from merging of NSs, makes the once but long ago
theoretical dream of understanding how NSs are composed, a
scientifically strong and overwhelming reality. For a mainstream of
textbooks and reviews on NSs, we refer the reader, for example,  to Refs.
\cite{Haensel:2007yy,Friedman:2013xza,Baym:2017whm,Lattimer:2004pg,Olmo:2019flu}.

Even after so much time since the first pulsar was observed far
back in 1967 by Jocelyn Bell, to date the crust and core of NSs
are still mysteries for scientists. The reason is that NSs consist
mainly of nuclear matter, the density of which nearly after the
crust-core interface becomes supranuclear, and at the outer core
bottom is nearly twice the nuclear saturation density, which is
approximately $\rho_0\sim 2.8\times 10^{14}$g$/\mathrm{cm}^3$.
Recall that the saturation density of nuclear matter is determined
by the masses and radii of stable nuclei, and a simplistic value
can be calculated by assuming spherical nuclei of mass number $A$
in total. The behavior of nuclear matter at supranuclear densities
is to date one of the greatest mysteries of theoretical
astrophysics and nuclear physics of course. The main problem is
that such high densities cannot be reached by experiments
conducted on Earth, thus, for the moment, in most cases,
theoretical astrophysicists rely on non-relativistic (mostly)
results obtained by nucleon-nucleon scattering, or by strongly
coupled Coulomb systems experiments. Moreover, to date, one can
have strong hints about the main characteristics of the equation
of state (EoS) for the outer core, which ends at densities of the
order $\rho\sim 2\rho_0$, where $\rho_0$ is the nucleus energy
density $\rho_0\sim 2.8\times 10^{14}$$g/cm^3$, but, at higher
densities, there are only speculations for the state of  matter,
with the possibilities being that the EoS of matter is determined
by ordinary symmetric nuclear matter, hyperons, strange quarks, or
the EoS is determined by the condensation of Kaons or even of
Pions. Novel approaches exist, based on firm phenomenological
data, that may provide a piecewise polytropic EoS, valid up to
large central densities \cite{Read:2008iy,Read:2009yp}.

Apart from nuclear matter, all the other possibilities we
mentioned above, belong to exotic classes of NS models. Besides
the standard General Relativity (GR) description of NS, there is
also the possibility of Extended Gravity
\cite{reviews1,reviews2,reviews3,reviews4,reviews5,reviews6,dimo,
book} that may also describe NSs. In the context of Extended
Gravity, large mass neutron stars can be harbored easily, with
masses inside the mass gap region, and one is able to even solve
mysteries of the standard GR approach, such as the hyperon puzzle
\cite{Astashenok:2014pua}. Let us note here that by mass gap
region, we mean the gap between the observationally confirmed
lowest mass limit of astrophysical black holes, and the maximum
neutron star masses, with the mass gap region being $M\sim 2.5-5
M_{\odot}$. One may claim that modified gravity is not standard
gravity, especially in the Jordan frame, and that such
descriptions are new physics. But this claim may be refuted due to
the fact that the same claim could be done for the Newtonian
description of stars, and the GR one. If one sticks on the
foundations of Newtonian gravity, then the data indicate that NSs
cannot be appropriately described by Newtonian gravity. On the
other hand, GR, for which the Newtonian theory is the weak field
limit, perfectly describes medium mass NSs. Thus a generalization
of Newtonian gravity, namely GR, is what actually perfectly
describes medium mass NS. Therefore, a generalization of GR,
namely Extended Gravity, may be a more appropriate theory that can
accurately describe large mass NSs, and accurately predict their
maximum mass limit for a wide range of EoS. Hence, it is possible
that Extended Gravity is not new physics, but the actual physical
description of large mass NSs, with the latter being generated for
smaller central densities or even equal, compared to GR. On the
other hand, Extended Gravity, at least in its curvature based
formulation,  can be seen as the natural extension of GR as soon
as one wants to address utraviolet  and infrared  phenomena which
escape the standard description. In the first case, the lack of a
self-consistent theory of Quantum Gravity needs the semiclassical
approach of quantum field theory formulated on curved space-time
\cite{Birrell}. In the latter case, the big puzzle of dark side,
not yet solved at experimental level, can be encompassed in the
framework of gravitational phenomena \cite{Mauro,Curv}.

In this paper, we shall investigate the causal limit maximum mass
of NSs in the context of $f(R)$ gravity formulated in the Jordan
frame. We shall adopt the stiffest EoS one can use,
the causal limit EoS, with variable speed of sound. In this context,
 we shall investigate the causal limit of NS maximum mass in
the $f(R)$ frame, and how this limit is affected by the sound speed.
 The causal limit EoS will be linked at densities of the
order $\sim 2 \rho_0$ to the SLy EoS \cite{Douchin:2001sv}, with
$\rho_0$ being the saturation density, which is standard in the
literature. However, we shall also include  the case
where the transition density is $\rho_t\sim  \rho_0$, mainly
motivated by the fact that the exact nuclear matter behavior
is known for densities of the order of  saturation density.

The results of our investigation will be critically examined
discussing the pros and cons of  Extended Gravity description of
NSs with masses in the mass gap region. Finally, we shall
critically take into account the possibility that the secondary
compact stellar object of GW190814 binary event
\cite{Abbott:2020khf} is a strange star. Although such a
possibility is exciting and can be well fitted by the
observational data \cite{Roupas:2020nua}, we shall discuss why a
NS, or even a black hole (BH) possibility are more compatible with
an Occam's razor approach for the secondary component of the
GW190814 event. The probability that a NS or a small mass BH are
candidates for the secondary component of GW190814, is also
stressed in the literature \cite{Biswas:2020xna,Bombaci}, where a
thorough analysis of data is performed.

The layout of the paper is the following. Sec II is devoted to a discussion of causal limit of maximum NS mass in the Jordan frame of $f(R)$ gravity. We consider  also  the role of Extended Gravity with respect to the gravitational wave event  GW190814.
The strange star perspective,  according to the data of GW190814 event, is discussed in Sec.III. Conclusions and perspectives are drawn in Sec. IV.

\section{Causal Limit of Maximum Neutron Star Mass in Jordan frame $f(R)$ Gravity and the Role of Extended Gravity for GW190814 Event}

The maximum mass allowed in a NS is one of the most difficult
 issues in NS physics, and it is strongly related with
the EoS and the adopted theory of gravity.  In other words,
gravity and the EoS determine the maximum NS mass. For standard
GR, the maximum mass obtained for the softest BPAL12 EoS is
$M_{max}\sim 1.4 M_{\odot}$ while for the stiffest, BGN2 EoS,  the
maximum mass is $M\sim 2.5 M_{\odot}$ \cite{Haensel:2007yy}.
Actually, for $Ne\mu$ cores, $M_{max}$ varies in the range $1.8 -
2.2M_{\odot}$, and, in presence of hyperons (assuming two body
interactions), the above limit becomes roughly $1.5 -
1.8M_{\odot}$. The existence of hyperons softens significantly the
EoS, unless a three body interaction is assumed for hyperons. In
this case,  the EoS may result significantly stiffen. Iff the
secondary component of GW190814 proves to be a NS, such a massive
NS, in the context of GR, indicates that the core should be
composed by nuclear matter solely. This issue however may not be
necessarily true in the context of modified gravity, since hyperon
based EoS also yield large masses, with the EoS not being too
stiff \cite{Astashenok:2014pua}, thus the hyperon puzzle can be
solved in the context of modified gravity.
\begin{figure}[h!]
    \centering
    \includegraphics[scale=0.35]{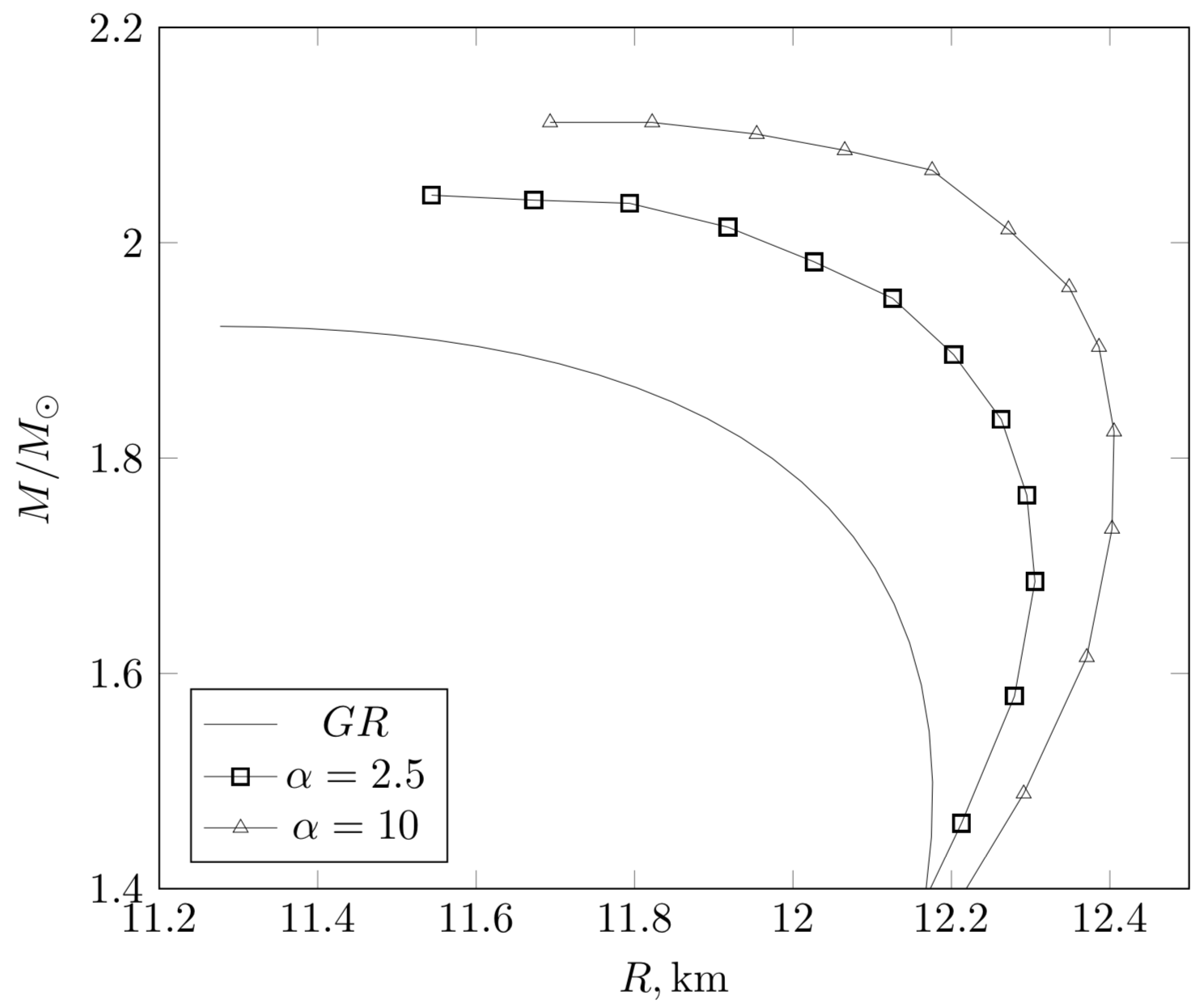}\includegraphics[scale=0.35]{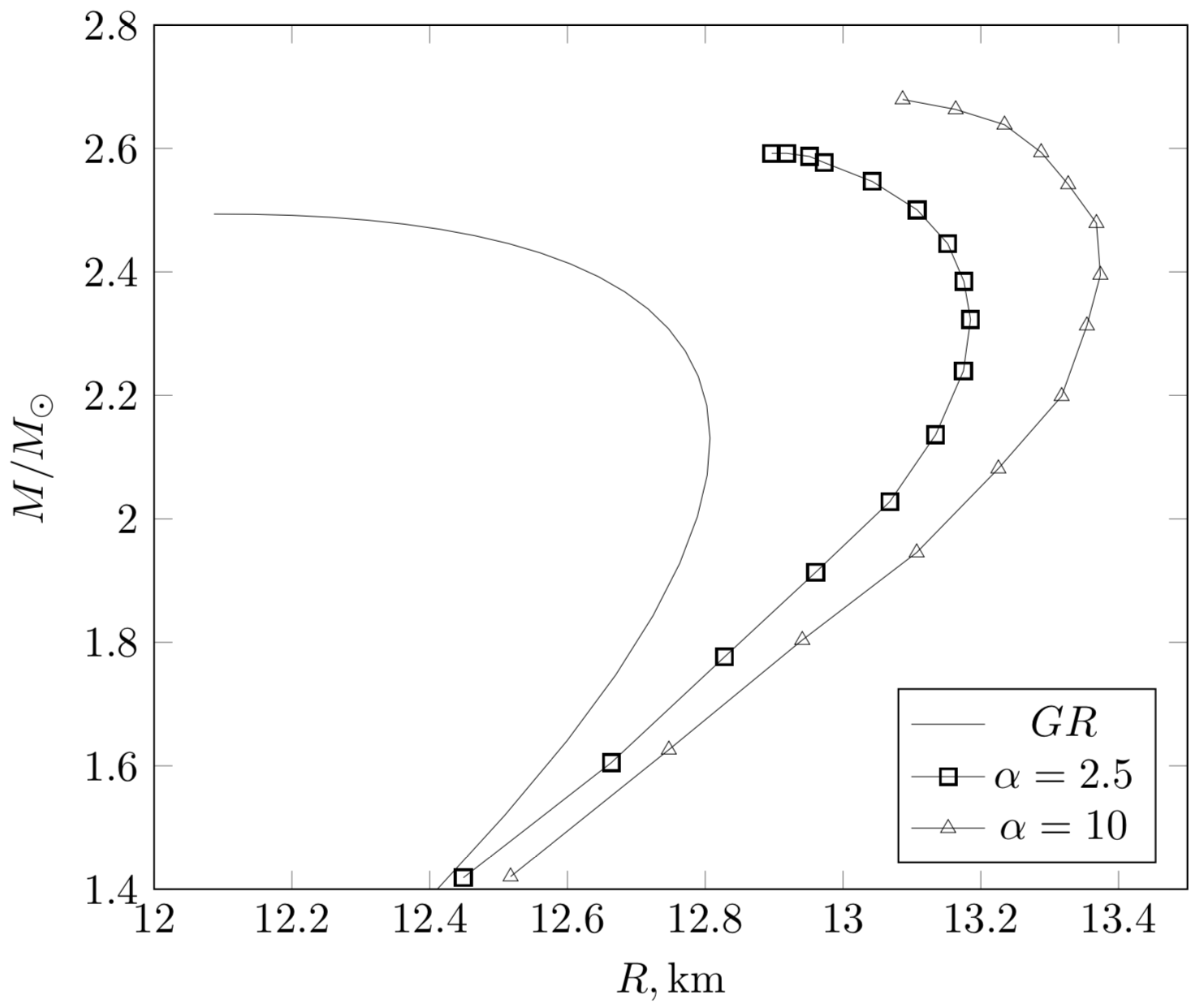}\\
    \includegraphics[scale=0.35]{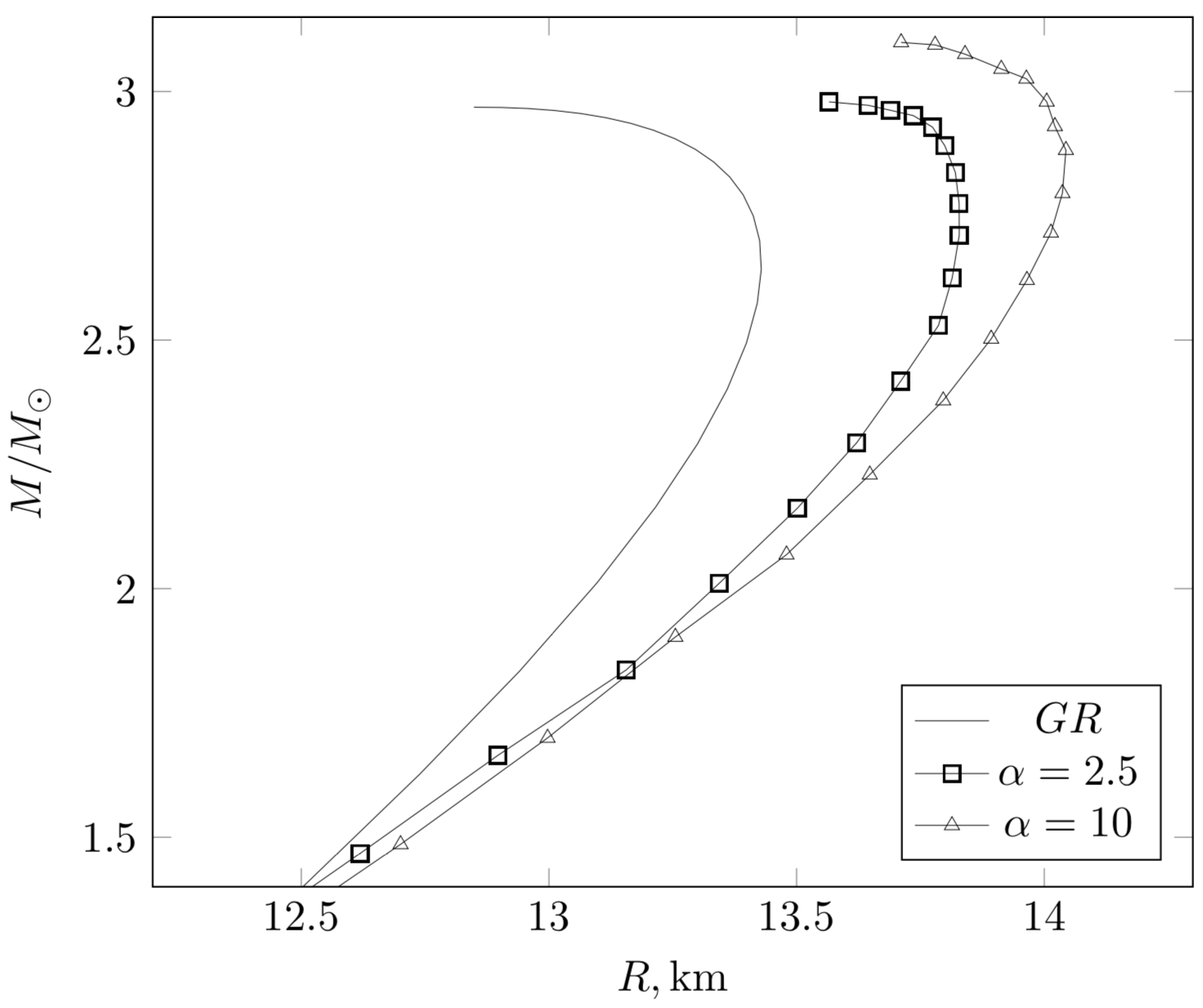}
    \caption{Mass-radius relation for equation of state SLy+(5) with $\rho_u=2\rho_0$ at $v^2=c^2/3$ (upper right panel), $v_s^2=3c^2/5$ (upper left panel), $v_s^2=c^2$ (down panel).}
    \label{plot1}
\end{figure}
The very own existence of a maximum NS mass is due to
the combined effect of GR (or Extended Gravity) and of EoS of
supranuclear densities cold nuclear matter. GR calculations
indicate that the maximum NS mass is mainly determined by the
EoS at supranuclear densities of at least $\rho\geq \rho_0$
\cite{Haensel:2007yy}, with $\rho_0$ being the saturation density.
Causality is a severe constraint for most EoSs, so a reliable EoS
must respect causality. The EoS can become superluminal when
$dP/d\rho>1$ in natural units, and also can become ultrabaric when
$P>\rho$.

The question is whether an EoS which predicts
superluminality in the speed of sound, is inconsistent with Lorenz
invariance and causality. The answer is no, however for the most
known and frequently used EoSs, like the SLy, superluminality
occurs for energy densities where the NS becomes
hydrodynamically unstable \cite{Haensel:2007yy}. The same applies
even for the stiffest BGN2 EoS.

Taking into account the stability
of nuclear matter at high densities ($\frac{dP}{d\rho}>0$), and
the subluminality condition ($\frac{dP}{d\rho}\leq 1$ in natural
units), the widely accepted causal upper bound limit on the
maximum  mass for the static NS case is
\cite{Rhoades:1974fn,Kalogera:1996ci},
\begin{equation}\label{causalupperbound}
M_{max}^{CL}=3M_{\odot}\sqrt{\frac{5\times
10^{14}g/cm^{3}}{\rho_u}}\, ,
\end{equation}
where $\rho_u$ is the maximum density for which the EoS is well
known, with corresponding pressure  being $P_u(\rho_u)$. Also the
causal limit maximum mass of Eq. (\ref{causalupperbound}) is
obtained by assuming the causal limit EoS,
\begin{equation}\label{causallimiteos}
P_{sn}(\rho)=P_{u}(\rho_u)+(\rho-\rho_u)c^2\, .
\end{equation}
A safe statement, also widely known in astrophysics, is that the
actual maximum mass of a static NS ($P\succeq 3$ms), built from
baryon mass, is
\begin{equation}\label{3solarmasslimit}
M_{max}\leq 3M_{\odot}\, .
\end{equation}
Rotation increases the causal upper limit of the maximum NS mass
to,
\begin{equation}\label{causalrot}
M^{CL,rot}_{max}=3.89M_{\odot}\sqrt{\frac{5\times
10^{14}g/cm^{3}}{\rho_u}}\, .
\end{equation}
The maximum mass of a NS in Extended Gravity  is the most
serious issue to be addressed. Some useful insights may be
provided by examining the causal limit maximum mass for NSs in the
context of Extended Gravity. This limit will provide us with an
upper limit on the maximum mass, thus it is an indication of what
is the most extreme upper bound on NS masses  in the context of
Extended Gravity.

Here we shall calculate numerically
the causal limit of the maximum mass for NSs in the context of
 $f(R)$ gravity considering the Jordan frame. More importantly: Is  there a gap
between maximum NS mass and lowest BH mass, or can they  be
continuously found even in the mass gap region $M\sim 2.5- 5
M_{\odot}$ \cite{Bailyn:1997xt}? In the literature, some ways for
discriminating NS and BH are given \cite{Fasano:2020eum}. These
criteria are based on gravitational waves tidal deformability
measurements.
\\
\emph{However, even if one is able to
discriminate a BH and a NS in binaries, the question is, what is
the theoretical upper limit for NS masses and the lower limit of
BH masses (if any)? The true question towards understanding the
mass gap region is, what is the maximum allowed mass of a NS.}
\\
Hence, this upper limit on  NS mass for Extended Gravity is the
one obtained by using a causal EoS, which we shall calculate
numerically. The questions posed above are important and closely
related to future LIGO-Virgo observations and the upcoming LISA
project. Thus NSs are truly modern laboratories that may
put to test several theoretical physics frameworks, such as
modified gravity in its various forms, both in the Jordan
\cite{Astashenok:2020qds,Capozziello:2015yza,Astashenok:2014nua,Astashenok:2013vza,Laurentis}
or Einstein frame
\cite{Pani:2014jra,Ramazanoglu:2016kul,Yazadjiev:2014cza}, or even
string-motivated gravity
\cite{Berti:2020kgk,Silva:2017uqg,Pani:2011xm}, theoretical
astrophysics, nuclear physics \cite{Kanakis-Pegios:2020kzp} and
even particle physics
\cite{Safdi:2018oeu,Astashenok:2020cfv,Day:2019bbh,Edwards:2020afl,Brito:2017zvb}.
\begin{figure}[h!]
    \centering
    \includegraphics[scale=0.35]{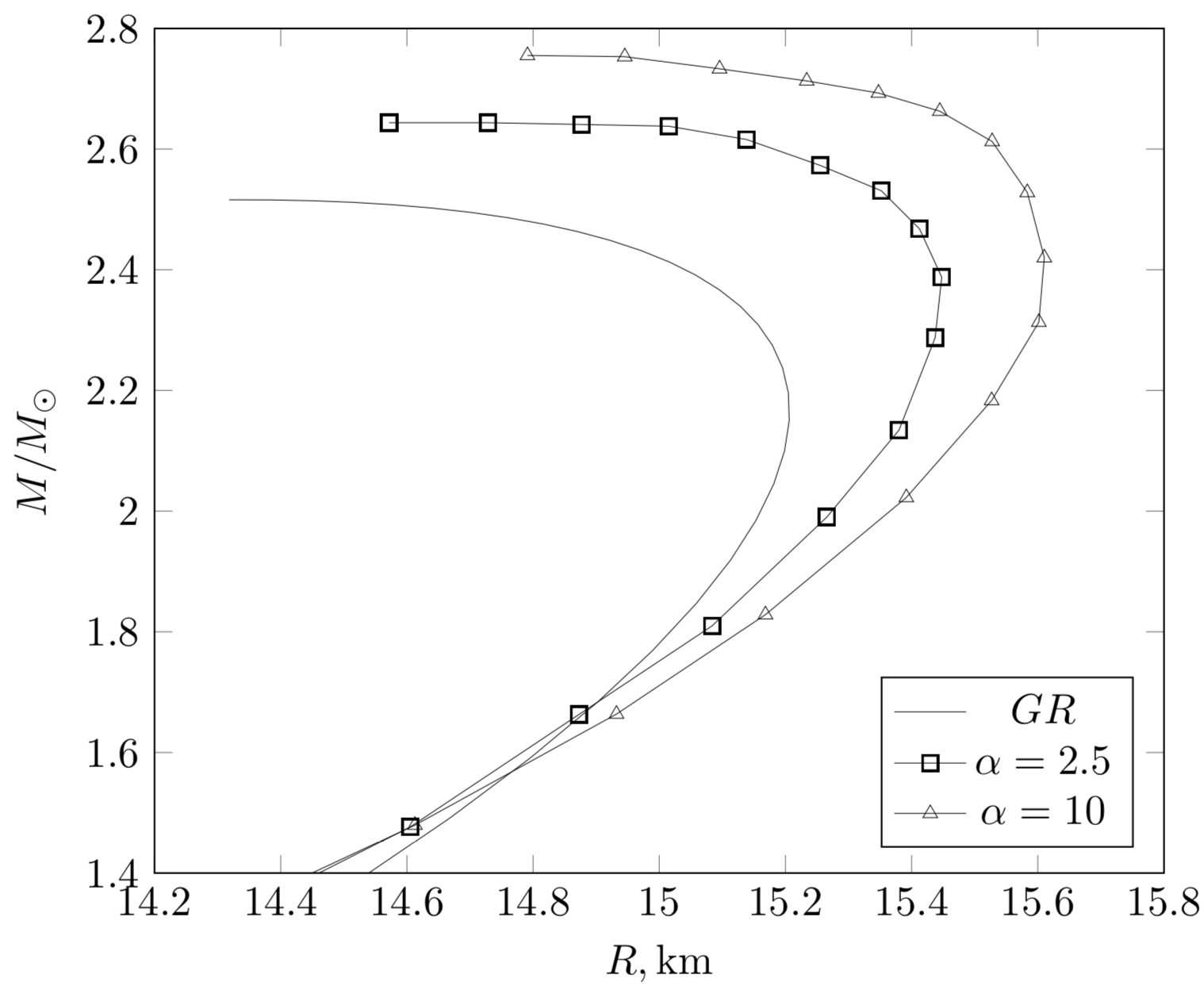}\includegraphics[scale=0.35]{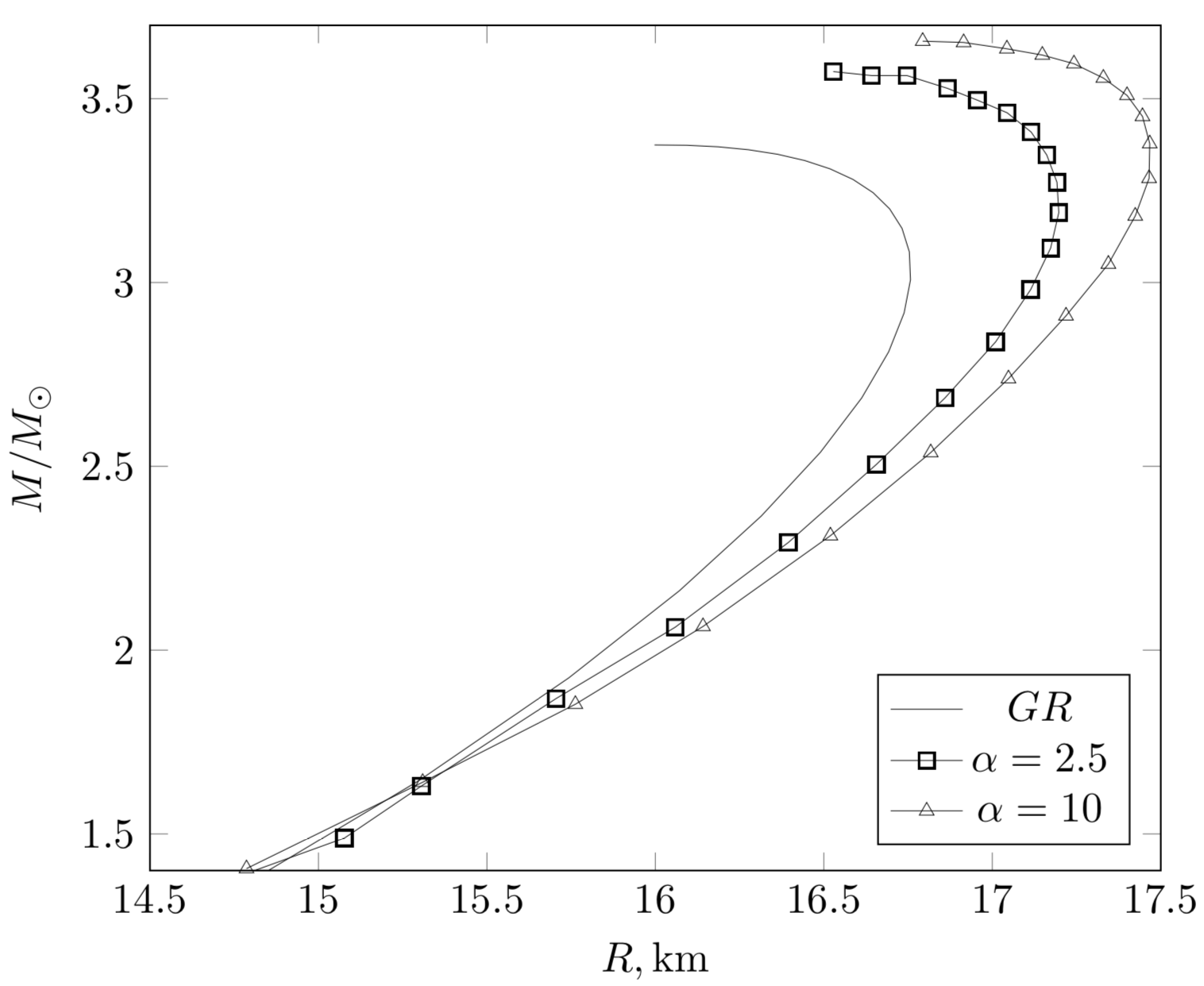}\\
    \includegraphics[scale=0.35]{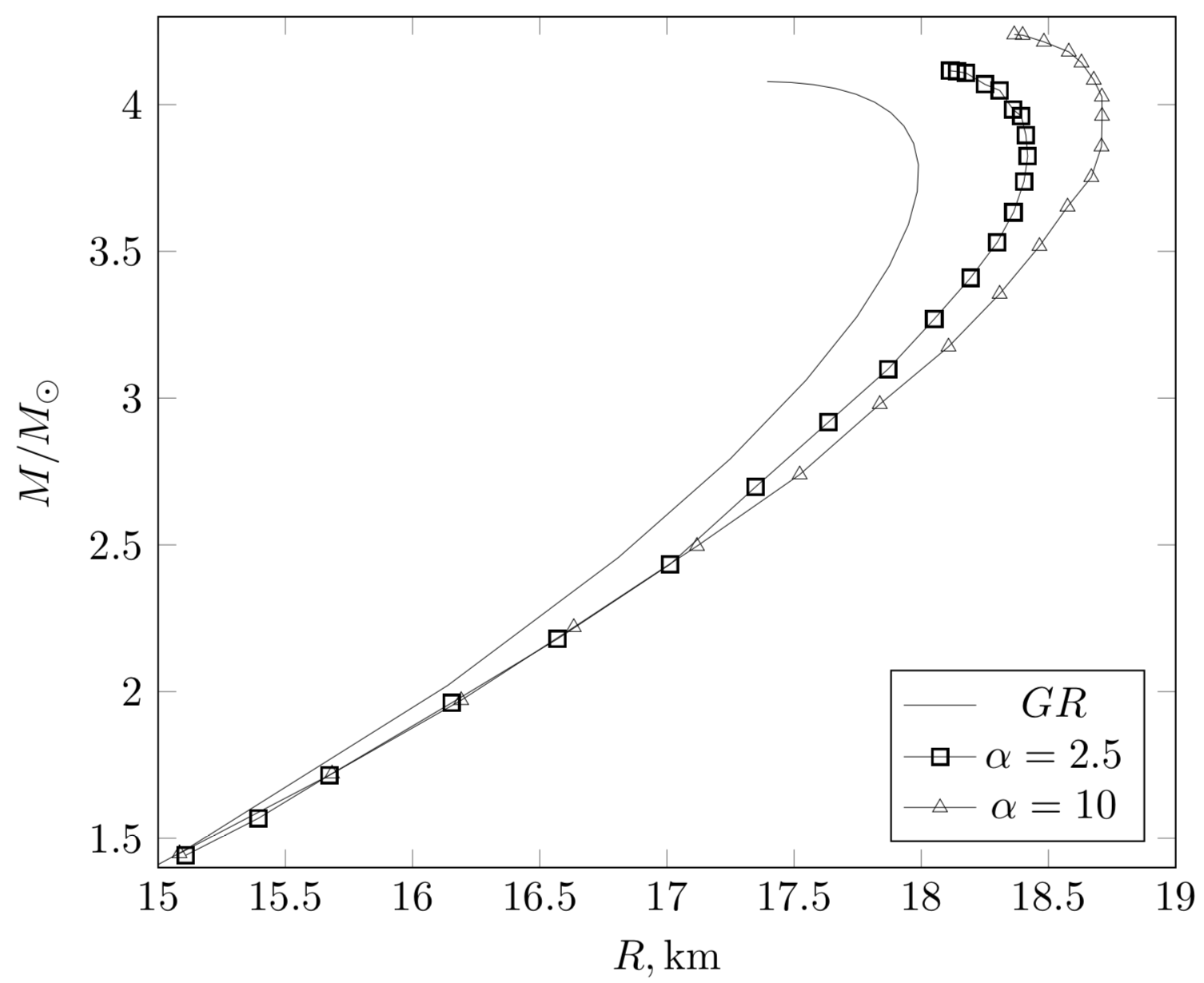}
    \caption{Mass-radius relation for equation of state SLy+(5) with $\rho_u=\rho_0$ at $v_s^2=c^2/3$ (upper right panel), $v_s^2=3c^2/5$ (upper left panel), $v_s^2=c^2$ (down panel).}
    \label{plot2}
\end{figure}
The calculation of the maximum mass causality limit or
subclasses of scalar-tensor gravity has been performed in
\cite{Sotani:2017pfj} and the results are rather interesting. In
this paper, we shall perform the same calculation in the context
of the Jordan frame of $f(R)$ gravity, by using the causal limit EoS of
Eq. (\ref{causallimiteos}), with a variable speed of sound $v_s$.
We shall adopt the notation of \cite{Sotani:2017pfj} for our
study. The causal limit EoS we shall use in this paper is,
\begin{equation}\label{causallimiteosnew}
P_{sn}(\rho)=P_{u}(\rho_u)+(\rho-\rho_u)v_s^2\, ,
\end{equation}
and although there exist a conjecture that the sound speed
should be less than $c/\sqrt{3}$ \cite{Sotani:2017pfj}, we shall
assume that the sound speed  is in the range $c^2/3\leq
v_s^2\leq c^2$, as in Ref. \cite{Sotani:2017pfj}. Also $P_u$ and
$\rho_u$ in Eq. (\ref{causallimiteosnew}) correspond to the
pressure and density of the well known segment of a low-density
EoS, with $\rho_u\simeq 2\rho_0$. We shall refer to these
quantities as transition pressure and transition density
hereafter, and we shall assume that the low-density EoS is the SLy
\cite{Douchin:2001sv} EoS. Thus, $\rho_u=2.7223\times 10^{14}$
$g/cm^{3}$, and the corresponding pressure is $P_u=3.5927\times
10^{33}$ $erg/cm^{3}$. But motivated by the fact that ground based
nuclear experiments truly describe well nuclear matter up to the
saturation density, we shall perform the same numerical analysis
for the SLy EoS, by taking the transition density to be of the
order of the saturation density, namely $\rho_u\sim \rho_0$.

In the
following ,we shall solve the $f(R)$ gravity
Tolman-Oppenheimer-Volkoff (TOV) equations numerically for the EoS
 given in Eq. (\ref{causallimiteosnew}) assuming
various values of the sound speed in the range $c^2/3\leq
v_s^2\leq c^2$.

In order to be self-contained, let us briefly
recall the $f(R)$ gravity framework for a spherically symmetric
compact object, and the corresponding TOV equations. The $f(R)$
action is given by,
\begin{equation}\label{action}
{\cal A}=\frac{c^4}{16\pi G}\int d^4x \sqrt{-g}\left[f(R) + {\cal
L}_{{\rm matter}}\right]\,,
\end{equation}
where $g$ denotes the determinant of the metric tensor
$g_{\mu\nu}$ and ${\cal L}_{\rm matter}$ is the perfect fluid
matter Lagrangian. Upon varying (\ref{action}) with respect to the
metric $g_{\mu\nu}$, we obtain the field equations  \cite{reviews4},
\begin{equation}
\frac{df(R)}{d R}R_{\mu\nu}-\frac{1}{2}f(R)
g_{\mu\nu}-\left[\nabla_{\mu} \nabla_{\nu} - g_{\mu\nu}
\Box\right]\frac{df(R)}{dR}=\frac{8\pi G}{c^{4}} T_{\mu \nu },
\label{field_eq}
\end{equation}
where $\displaystyle{T_{\mu\nu}=
\frac{-2}{\sqrt{-g}}\frac{\delta\left(\sqrt{-g}{\cal
L}_m\right)}{\delta g^{\mu\nu}}}$ denotes the energy-momentum
tensor of the matter perfect fluids.  The metric describing
spherically symmetric spacetimes is,
\begin{equation}
    ds^2= e^{2\psi}c^2 dt^2 -e^{2\lambda}dr^2 -r^2 (d\theta^2 +\sin^2\theta d\phi^2),
    \label{metric}
\end{equation}
where $\psi$ and $\lambda$ are $r$-dependent functions, with $r$
denoting the  radial coordinate. For the perfect
matter fluid  describing the   compact stellar interior, the energy-momentum tensor is
$T_{\mu\nu}=\mbox{diag}(e^{2\psi}\rho c^{2}, e^{2\lambda}p, r^2 p,
r^{2}p\sin^{2}\theta)$ where $\rho$ is the matter density and $p$
denotes the pressure \cite{weinberg}. The equations for the
stellar compact object are obtained by adding the hydrostatic
equilibrium condition from the contracted
Bianchi identities,
\begin{equation} \nabla^{\mu}T_{\mu\nu}=0\,,
\label{bianchi}
\end{equation}
which yield  the Euler conservation equation
\begin{equation}\label{hydro}
    \frac{dp}{dr}=-(\rho
    +p)\frac{d\psi}{dr}\,.
\end{equation}
Combining the metric \eqref{metric} and the field  Eqs.
(\ref{field_eq}), the equations describing the functions $\lambda$
and $\psi$ are \cite{capquark},
\begin{eqnarray}
\label{dlambda_dr} \frac{d\lambda}{dr}&=&\frac{ e^{2 \lambda
}[r^2(16 \pi \rho + f(R))-f'(R)(r^2 R+2)]+2R_{r}^2 f'''(R)r^2+2r
f''(R)[r R_{r,r} +2R_{r}]+2 f'(R)}{2r \left[2 f'(R)+r R_{r}
f''(R)\right]},
\end{eqnarray}
\begin{eqnarray}\label{psi1}
\frac{d\psi}{dr}&=&\frac{ e^{2 \lambda }[r^2(16 \pi p -f(R))+
f'(R)(r^2 R+2)]-2(2rf''(R)R_{r}+ f'(R))}{2r \left[2 f'(R)+r R_{r}
f''(R)\right]}\, ,
\end{eqnarray}
where the prime in both Eqs. (\ref{dlambda_dr}) and
(\ref{psi1}) denotes differentiation with respect to  the Ricci
scalar $R(r)$.

The above equations constitute the modified TOV equations, which,
for $f(R)=R$, reduce to the standard GR TOV equations
\cite{rezzollazan,landaufluid}. In the context of
$f(R)$ gravity, the Ricci scalar is dynamically evolving with
respect to the variable $r$, thus the TOV equations should be
solved together with the following equation
\begin{equation}\label{TOVR}
\frac{d^2R}{dr^2}=R_{r}\left(\lambda_{r}+\frac{1}{r}\right)+\frac{f'(R)}{f''(R)}\left[\frac{1}{r}\left(3\psi_{r}-\lambda_{r}+\frac{2}{r}\right)-
e^{2 \lambda }\left(\frac{R}{2} + \frac{2}{r^2}\right)\right]-
\frac{R_{r}^2f'''(R)}{f''(R)},
\end{equation}
which is derived from the trace of Eqs.\eqref{field_eq} by using
the metric \eqref{metric}.

\begin{table}[htbp]%[H]
%\begin{centering}
\begin{tabular}{|c|c|c|c|c|c|}
  \hline
  EoS  & $\alpha$,         & $M_{max}$,    & $R_{s}$, & $\Delta M_{max},$  \\
       & $r_{g}^2$    & $M_{\odot}$   & km     &  $M_{\odot}$          \\
    \hline
    \multicolumn{5}{|c|}{$\rho_u=2\rho_0$}\\\hline
                    & 0  & 1.92 & 11.28  & 0      \\
                    & 0.25 & 1.97 & 11.45 & 0.05 \\
       SLy+(5)      & 2.50  & 2.04 & 11.54  & 0.12   \\
 with $v_s^2=c^2/3$ & 10  & 2.11 & 11.69 & 0.19    \\
        \hline
                    & 0  & 2.49 & 12.09  & 0      \\
                    & 0.25 & 2.53 & 12.45 & 0.04    \\
    SLy+(5)          & 2.50  & 2.59 & 12.90  & 0.10   \\
  with $v_s^2=3c^2/5$ & 10  & 2.68 & 13.09 & 0.19    \\
        \hline
                    & 0  & 2.97 & 12.85  & 0      \\
                    & 0.25 & 2.93 & 13.28 & -0.04    \\
    SLy+(5)          & 2.50  & 2.98 & 13.57  & 0.01   \\
  with $v_s^2=c^2$ & 10  & 3.10 & 13.71 & 0.13    \\
  \hline
  \multicolumn{5}{|c|}{$\rho_u=\rho_0$}\\\hline
                    & 0  & 2.52 & 14.35  & 0      \\
                    & 0.25 & 2.58 & 14.45 & 0.06    \\
SLy+(5)             & 2.50  & 2.65 & 14.57  & 0.15   \\
 with $v_s^2=c^2/3$ & 10  & 2.76 & 14.79 & 0.24    \\
        \hline
                    & 0  & 3.37 & 16.05  & 0      \\
                    & 0.25 & 3.44 & 16.32 & 0.07    \\
  SLy+(5)           & 2.50  & 3.57 & 16.53  & 0.20   \\
  with $v_s^2=3c^2/5$ & 10  & 3.66 & 16.78 & 0.29    \\
        \hline
                    & 0  & 4.08 & 17.39  & 0      \\
                    & 0.25 & 4.05 & 17.90 & -0.03    \\
  SLy+(5)           & 2.50  & 4.12 & 18.11  & 0.04   \\
  with $v_s^2=c^2$ & 10  & 4.23 & 18.37 & 0.15    \\
  \hline

\end{tabular}
\caption{Parameters of compact stars (maximum mass for given
EoS and corresponding radius) for various equation
of states in GR (i.e., $\alpha=0$) and for some
values of $\alpha$ in $f(R)=R+\alpha R^2$ gravity.
Parameter $\alpha$ is given in units of $r_g^2=4G^2M^2_\odot/c^4$
where $r_g$ is gravitational radius of Sun. It is $\Delta M_{max}=
M_{max}-M_{max}^{GR}$} \label{Table1}
\end{table}

Thus, in the following we shall solve the TOV Eqs.
(\ref{hydro}), (\ref{dlambda_dr}) and (\ref{psi1}) together
with (\ref{TOVR}), for
\begin{equation}\label{frdef}
f(R)=R+\alpha R^2\, ,
\end{equation}
where the parameter $\alpha$ is expressed in units of
$r_g^2=4G^2M^2_\odot/c^4$ where $r_g$ is the gravitational radius of
Sun. The nuclear matter inside the stellar object will be assumed
to satisfy the EoS (\ref{causallimiteosnew}). The sound speed is assumed to be a free parameter that may vary, and
also the transition density in Eq. (\ref{causallimiteosnew}) is
be chosen to be that of the SLy EoS at $\rho\sim 2\rho_0$, which
 is $\rho_u=2.7223\times 10^{14}$ $g/cm^{3}$.
$P_u$ is the corresponding pressure which is
$P_u=3.5927\times 10^{33}$ $erg/cm^{3}$ \cite{Douchin:2001sv}. The
same analysis can be performed by assuming  $\rho_u\sim
\rho_0$.

The results of  numerical analysis are presented in
Table \ref{Table1} and in Figs. \ref{plot1} and \ref{plot2}. In
Fig. \ref{plot1},  we present the mass-radius relation for the
causal EoS of Eq. (\ref{causallimiteosnew}) with the low density
EoS at $\rho_u=2\rho_0$  being the SLy. The upper left plot
corresponds to $v_s^2=c^2/3$, the upper right plot to
$v_s^2=3c^2/5$ and the bottom plot to  $v_s^2=c^2$. Also the same
numerical analysis for the transition density  $\rho_u=\rho_0$ is
presented in Fig. \ref{plot2}. The resulting picture is rather
interesting and rich in qualitative conclusions that can be made.
Specifically, it seems that, for all the values of the parameter
$\alpha$, when the sound speed  is less than
that of light, the causal limit maximum NS mass is larger
in comparison to the causal limit maximum mass of  GR case, for
both the cases $\rho_u=2\rho_0$ and $\rho_u=\rho_0$. However, when
$v_s=c$, for small values of the parameter $\alpha$, the GR causal
maximum mass is larger if compared to the one of  $R^2$ model, but, for larger
values of the parameter $\alpha$, the $R^2$ model overwhelms over
the GR values. For the standard values in the literature, for the
transition density $\rho_u=2\rho_0$, the results on causal limit
maximum mass indicate that NSs for the specific $f(R)$ model and
the low density SLy EoS, the masses belong to the mass gap region,
but not as deeply as we expected. In fact the GR results are quite
close, so even in the context of the $R^2$ gravity, the $\sim 3
M_{\odot}$ upper mass barrier (\ref{3solarmasslimit}) seems to
remain the same. Finally, as a last comment, it is worth noticing that
when the transition density of the causal EoS is assumed to be
equal to the nuclear saturation density, the maximum masses are
larger if compared to the case when  the transition density is twice
the saturation density.

\section{The Strange Star Perspective: Is It  Possible  or Probable According to GW190814 Event?}

Using the Occam's razor approach, the answer to the question posed
in the title is possible for the moment, but
not probable. Although nature will continue to surprise us, and
the strange star perspective is very exciting for theoretical
physics, for the moment we evince that the strange star is just a
possibility, but not a probability according to the description of the
GW190814 event.

The most important argument that makes the  strange star
possibility small for the GW190814  event is the very  well accepted procedure of how stars,
 undergoing thermonuclear evolution, remain hydrodynamically
stable. In ordinary stars, and NSs in particular, gravity keeps the
star in hydrodynamical equilibrium, it keeps the star compact.
Gravity counteracts the Fermi pressure of the degenerate fermionic
material in ordinary stars. In addition, for
NSs, gravity compensates even the repulsive
neutron-neutron interaction occurring well beyond the superfluid
crust and outer core, at supranuclear densities. Thus gravity and
Fermi pressure are the forces that keep an ordinary star and NS, in particular,
in hydrodynamical equilibrium.

On the other hand, strange stars  rely on the synergy of gravity
and Quantum Chromodynamics (QCD), to compensate the Fermi pressure
of the hadronic matter. In the strange stars, the ground
state of hadronic matter is, of course, self-bound strange quark
matter. Thus, this argument is the strongest argument against
strange quark stars.

In order for a strange star to be the secondary object of the
GW190814 event, one must change the very own well established fact
that gravity is the only opposing force that compensates the Fermi
pressure of nuclear matter in stars, while in the case of strange
stars, QCD and gravity compensate the Fermi pressure. Thus NSs, in
contrast to atomic nuclei and strange stars, are solely bound by
gravity, hence they serve as the Occam's razor description for the
secondary component of the binary merging GW190814.

To our knowledge, a smoking gun evidence for
strange stars would be either the observation of an apparent
radius $R_{\infty}=\frac{R}{\sqrt{1-\frac{r_g}{R}}}$  below 9Km,
that is $R_{\infty}<9$km, or the observation of half-millisecond
pulsars. In principle, the sub-millisecond periods cannot be
sustained by ordinary nuclear matter NSs, since this high spin
rotation would severely affect its hydrodynamic stability. Strange
stars however can sustain such spin periods since they are self-bound by QCD and  gravity.

Apart from the above smoking gun evidence, there
is no other possible and scientifically consistent way to confirm the
existence of strange stars. Strange stars may have been formed
primordially, during the early Universe, and of course during the
so-called \textit{Asymptopia era}. There were two possibilities, that may
work in favor or against the strange star hypothesis, either that
the \textit{primordial strangelets} \cite{Friedman:1990qz} survived after
the Asymptopia.  If they did so, no NS presence would be
justified in the Universe, and this argument would work against
the strange stars. If on the other hand some primordial
strangelets \cite{Friedman:1990qz} survived, then these could
co-exist with NSs and would not pollute the Universe with their presence
\cite{Haensel:2007yy}.

To our personal opinion, the primordial
strangelets fall into the same category where primordial BHs
belong, so both strangelets and primordial BHs are
primordial hypothetic objects, and the question is how did these
survive and why did NSs appear in the presence of primordial
strange stars in the first place? These questions cannot be
answered for the time being in a self-consistent way, a fact that
casts doubt on the same existence of strange stars. Still, no one can
exclude strange stars as possible compact stellar objects, but
these are not probable for the moment.

Another puzzle is the very own EoS of strange stars. One could
claim that it is QCD-derivable, where one assumes a
$\mathcal{O}(1)$GeV energy scale (or a few GeV) one gluon exchange
interaction among quarks (tree-order process), and thus the soft
coupling of QCD applies, and the EoS is easily found. However,
the Asymptopia state we just described, cannot be easily reached in
strange stars. In order for the soft QCD coupling physical state
to occur even at tree order (one gluon exchange), one must have
high quark densities, many orders above $\rho\sim 10^{15}g/cm^3$,
and energies $\sim 1$GeV. These conditions are not realized in
strange stars, and are only realized during the pre-inflationary
epoch where $SU(N)$ GUTs are unbroken or are about to be
broken. On the other hand, for NSs the compressibility modulus of
symmetric nuclear matter is higher or smaller than $K_0=220$MeV,
and this can be achieved by using stiff or soft EoSs. In all cases
though, the density is of the order of a few $\rho\sim
10^{15}g/cm^3$,  the two compact stellar objects are
 quite significantly different

Another argument which works to date against strange stars, and
makes them a hypothetical and not the Occam's razor choice, is the
existence of Pulsar glitches.  They are associated with
the undoubted presence of NS crust, and they occur due
to the sudden unpinning of superfluid crust vortices, situated
near the neutron dripping density $\rho_{ND}\sim 0.001 \rho_0$.
According to this interpretation of glitches, the superfluid
component of the crust must have at least $1\%$ of the total
momentum of inertia of the NSs. Evidence for the presence
of a crust in NS is mainly supported by the observed free
precession of radio pulsars. Strange stars, on the other hand,
cannot describe glitches, so this observation serves, at the moment,
against the presence of strange stars.

An interesting argument, but still questionable, is the absence of
r-process electromagnetic radiation during the  GW190814 merging process,
which could act against the BH-NS candidates. As it is known, BH
mergers are thought to produce no electromagnetic radiation, thus
one claim that the GW190814 is a BH merger due to the fact that no
kilonova observation was observed and secondly, it cannot be a
NS-BH merger because we have never observed such a binary. These
claims however are not strongly supported by all the theoretical
and observational facts.  Firstly, the GW190814
event was six times farther in comparison to the GW170817 event,
where a kilonova was also observed. Moreover, if the secondary
object of GW190814 was a NS, it would have been swallowed by the
$\sim 23M_{\odot}$ BH. Again, if the small mass component of the
GW190814 event was a NS, no electromagnetic or $r$-process ejecta
would have be observed because the tidal forces would be small
\cite{Fasano:2020eum}. In the same scenario, simulations indicate
that the disk mass would be small, so the electromagnetic
processes would have been difficult to detect
\cite{Fasano:2020eum,Foucart:2018rjc,Zappa:2019ntl,Foucart:2019bxj}.
Moreover, the observation or not of a kilonova depends on the line
of sight, and on the large uncertainties of the source, as these
are determined by LIGO and Virgo.

Furthermore, there is the argument that strong interactions play a
significant role both in strange star and in NS physics,
so both NSs and strange stars are likely to be found, since the
same physics of strong interactions control them. This is not
correct of course, since if we recall that the maximum mass of a
NS for a free Fermi neutron gas is $\sim0.72 M_{\odot}$, the
famous Oppenheimer-Volkoff limit, which by simply assuming beta
equilibrium becomes $\sim0.7 M_{\odot}$. However, the presence of
a strongly interacting fluid is verified by the existence of NSs
with masses quite larger than the Oppenheimer-Volkoff limit. Thus
strong interactions make their presence apparent in NSs too,
however in the most part of  NSs, in the outer core and inner
core, and always at supranuclear densities, where the repulsive
component of the neutron-neutron interaction takes place, and thus
NSs are not held together by strong interactions, but by gravity
itself.

%%%%%%%%%%%%%%%%%%%%%%%%%%%%%%%%%%%%%%%%%%%%%%%%%%%%%%%%%%%%%%%%%%%%%%%%%%%%%%%%%%%%%%%%%%%%%%%%%%%%%%%%%%%%%%%%%%%%%%%%%here...

\section{Concluding Remarks}

In this work, we investigated the causal maximum mass limit of NSs in
$f(R)$ gravity   focusing on the $R^2$
model. Particularly, we used the stiffest EoS one can use, the
causal EoS, and we numerically solved the TOV equations for the
$R^2$ model in order to find the maximum NS mass for the causal
EoS. With regard to the causal EoS, it consists of two parts, the
low density and the high density, and, for the low-density
part, we assumed it consists of the SLy EoS. The density where the
low and high density parts meet, is called the transition density,
denoted by $\rho_u$, and, for the transition density, we assumed that
it is equal to either twice the saturation density $\rho_0$ or it is
just equal to the transition density. Accordingly, we solved
numerically the TOV equations and  extracted the maximum causal
masses for various values of the free parameter of the model,
which is the coupling of the $R^2$ term denoted $\alpha$.  The
results are interesting since the conclusion is that, for
the choice $\rho_u=2\rho_0$, the maximum causal limit for the
$f(R)$ gravity is quite close to the GR maximum causal mass and
more importantly, the three solar masses causal limit is also
marginally respected by the $R^2$ model, for the SLy EoS at
least. In addition, a notable feature is that when the speed of
sound is equal to the speed of light, and simultaneously for small
values of the coupling parameter $\alpha$, the GR causal maximum
mass is larger compared to the $R^2$ model, and this feature was
also observed in a different context though in
\cite{Sotani:2017pfj}. However, this behavior does not hold true
for larger values of the coupling parameter $\alpha$. Moreover, it
is also notable that the maximum mass limits, for $\rho_u\sim
\rho_0$, are larger compared to the case $\rho_u\sim 2\rho_0$.

In view of the fact that the upper causal mass limit lies within
the mass-gap region (but not deeply as we verified, at least for
the $R^2$ model), and in relation with our previous work
\cite{Astashenok:2020qds}, the GW190814 event may be well
described by Extended Gravity. However another candidate
description for the secondary component of GW190814 is a strange
star, so we critically discussed the possibility that the large
mass component of the merging event GW190814 is a strange star.
As we evinced, by using the Occam's razor approach, this
possibility is deemed unlikely, at least for the moment. As it is
demonstrated in Ref. \cite{Biswas:2020xna}, the presence of a
large mass NS or a small mass BH is probable, not just possible.
Of course, Nature has a unique way to utterly change the
scientific perspective, but, for the moment, the most
logical way is to stick to the simplest solution that seems to
describe all compact astrophysical objects. That is, gravity and
Fermi pressure keep the stellar objects hydrodynamically stable,
not the synergy of QCD and gravity against Fermi pressure of
nuclear matter.

However, even if we stick with the NS and BH solutions, there are
still many issues that need to be appropriately addressed. The
most important is, what is the maximum mass allowed for NSs,
and what is the minimum mass for astrophysical BHs.
If NSs are found, with masses larger than $3M_{\odot}$, this
observation will clearly show that GR is unable to describe large
mass NSs, and thus Extended Gravity might be the optimal physical
description for NSs, combined with an appropriate EoS. Which EoS
though could be the optimal? The GW170817 event clearly excluded
several stiff EoS, like the WFF1 \cite{Bauswein:2017vtn}, however
the optimal, to our opinion, EoS are the SLy \cite{Douchin:2001sv}
the FPS EoS and piecewise polytropic EoSs
\cite{Read:2008iy,Read:2009yp}. SLy and FPS EoSs are the only EoSs
which provide a unified description of the crust core, since the
EoS is continuous at the crust-core interface. To our knowledge,
only these two EoS have this appealing property. But to our
opinion, the piecewise polytropic EoS, is a particularly appealing
description of matter at supranuclear densities, based on
optimization of data and simulation of polytropic EoSs
\cite{Read:2008iy,Read:2009yp}.

One may claim that any modified gravity it is not the standard  gravity,
especially the Jordan frame forms of modified gravity as
higher-order gravity.  But this claim is rather naive due to the
fact that the same claim could be done for the Newtonian
description of stars, and the general relativistic one. If one
sticks on the foundations of Newtonian gravity, then the data
indicate that NSs cannot be appropriately described by
Newtonian gravity. On the other hand, GR, for
which the Newtonian theory is the weak field limit, perfectly
describes medium mass NSs. Thus a generalization of
Newtonian gravity, namely GR, is what actually
perfectly describes medium mass neutron stars. Therefore, a
generalization of GR, namely Extended Gravity, may
be a more appropriate theory  accurately describing large
mass NSs, and accurately predicting their maximum mass
limit for a wide range of EoSs. In the case of modified gravity
description of NSs, gravity still describes the
hydrostatic equilibrium of NS, in contrast to the strange star
hypothesis, where the synergy of QCD and gravity keeps the star in
hydrodynamic equilibrium.

Now let us critically discuss what is the role of modified gravity
in the GW190814 description, adopting again the Occam's razor
approach. In the context of modified gravity, neutron stars are
kept in hydrodynamic equilibrium by gravity and Fermi pressure of
the supranuclear density matter. In the context of modified
gravity, and even scalar-tensor gravity, one may obtain larger
maximum neutron star masses for the same EoSs used in GR
\cite{Sotani:2017pfj}, as we also evinced in a  previous work
\cite{Astashenok:2020qds} and we also demonstrated in this work
for the causal limit EoS. Of course,  the causal EoS cannot describe
a physical NS, but it is an indication of the upper
limits of the maximum masses for a specific theory of gravity.
Also modified gravity descriptions of NS may  solve the
hyperon problem for NS masses larger than $2M_{\odot}$, due to the
fact that, in the context of GR, only stiff nuclear matter, without
hyperons, can achieve such large maximum NS masses
\cite{Astashenok:2014pua}. In modified gravity though, this is not
an issue, since large maximum NS masses can be achieved even for
hyperon related EoS \cite{Astashenok:2014pua}. A future
perspective task, that may shed some light on the question of what is
the maximum NS mass and which is the smallest BH mass, is the
study of the maximum baryon mass for static NSs. Specifically, the
maximum static baryon mass could serve as a lower limit for BHs,
since supermassive NSs with masses larger than the maximum baryon
masses will collapse to BHs. Thus studying the baryon masses for
Extended Theories of Gravity could be proven very useful for NS
physics, and we hope to address this issue in the near future.

\section*{Acknowledgments}

This work was supported by MINECO (Spain), project
PID2019-104397GB-I00 and PHAROS COST Action (CA16214) (SDO). SC acknowledges
 the support by  {\it Istituto Nazionale di Fisica Nucleare} (INFN) ({\it iniziative specifiche} MOONLIGHT2 and
 QGSKY). This work was partially supported by Ministry of Education of Russian Federation, project No FEWF-2020- 003.


\begin{thebibliography}{99}





\bibitem{Tolos:2020aln}
  L.~Tolos and L.~Fabbietti,
  %``Strangeness in Nuclei and Neutron Stars,''
  Prog.\ Part.\ Nucl.\ Phys.\  {\bf 112} (2020) 103770
  doi:10.1016/j.ppnp.2020.103770
  [arXiv:2002.09223 [nucl-ex]].


%\cite{Lattimer:2012nd}
\bibitem{Lattimer:2012nd}
J.~M.~Lattimer,
%``The nuclear equation of state and neutron star masses,''
Ann. Rev. Nucl. Part. Sci. \textbf{62} (2012), 485-515
doi:10.1146/annurev-nucl-102711-095018 [arXiv:1305.3510
[nucl-th]].
%599 citations counted in INSPIRE as of 24 Jan 2021

%\cite{Steiner:2011ft}
\bibitem{Steiner:2011ft}
A.~W.~Steiner and S.~Gandolfi,
%``Connecting Neutron Star Observations to Three-Body Forces in Neutron Matter and to the Nuclear Symmetry Energy,''
Phys. Rev. Lett. \textbf{108} (2012), 081102
doi:10.1103/PhysRevLett.108.081102 [arXiv:1110.4142 [nucl-th]].
%166 citations counted in INSPIRE as of 24 Jan 2021




%\cite{Horowitz:2005zb}
\bibitem{Horowitz:2005zb}
C.~J.~Horowitz, M.~A.~Perez-Garcia, D.~K.~Berry and
J.~Piekarewicz,
%``Dynamical response of the nuclear 'pasta' in neutron star crusts,''
Phys. Rev. C \textbf{72} (2005), 035801
doi:10.1103/PhysRevC.72.035801 [arXiv:nucl-th/0508044 [nucl-th]].
%124 citations counted in INSPIRE as of 24 Jan 2021


%\cite{Watanabe:2000rj}
\bibitem{Watanabe:2000rj}
G.~Watanabe, K.~Iida and K.~Sato,
%``Thermodynamic properties of nuclear 'pasta' in neutron star crusts,''
Nucl. Phys. A \textbf{676} (2000), 455-473 [erratum: Nucl. Phys. A
\textbf{726} (2003), 357-365] doi:10.1016/S0375-9474(00)00197-4
[arXiv:astro-ph/0001273 [astro-ph]].
%70 citations counted in INSPIRE as of 24 Jan 2021
%\cite{Sammarruca:2010gc}

%\cite{Shen:1998gq}
\bibitem{Shen:1998gq}
H.~Shen, H.~Toki, K.~Oyamatsu and K.~Sumiyoshi,
%``Relativistic equation of state of nuclear matter for supernova and neutron star,''
Nucl. Phys. A \textbf{637} (1998), 435-450
doi:10.1016/S0375-9474(98)00236-X [arXiv:nucl-th/9805035
[nucl-th]].
%684 citations counted in INSPIRE as of 24 Jan 2021




%\cite{Xu:2009vi}
\bibitem{Xu:2009vi}
J.~Xu, L.~W.~Chen, B.~A.~Li and H.~R.~Ma,
%``Nuclear constraints on properties of neutron star crusts,''
Astrophys. J. \textbf{697} (2009), 1549-1568
doi:10.1088/0004-637X/697/2/1549 [arXiv:0901.2309 [astro-ph.SR]].
%161 citations counted in INSPIRE as of 24 Jan 2021


%\cite{Hebeler:2013nza}
\bibitem{Hebeler:2013nza}
K.~Hebeler, J.~M.~Lattimer, C.~J.~Pethick and A.~Schwenk,
%``Equation of state and neutron star properties constrained by nuclear physics and observation,''
Astrophys. J. \textbf{773} (2013), 11
doi:10.1088/0004-637X/773/1/11 [arXiv:1303.4662 [astro-ph.SR]].
%385 citations counted in INSPIRE as of 24 Jan 2021


%\cite{Mendoza-Temis:2014mja}
\bibitem{Mendoza-Temis:2014mja}
J.~de Jes\'us Mendoza-Temis, M.~R.~Wu, G.~Mart\'\i{}nez-Pinedo,
K.~Langanke, A.~Bauswein and H.~T.~Janka,
%``Nuclear robustness of the r process in neutron-star mergers,''
Phys. Rev. C \textbf{92} (2015) no.5, 055805
doi:10.1103/PhysRevC.92.055805 [arXiv:1409.6135 [astro-ph.HE]].
%78 citations counted in INSPIRE as of 24 Jan 2021





%\cite{Ho:2014pta}
\bibitem{Ho:2014pta}
W.~C.~G.~Ho, K.~G.~Elshamouty, C.~O.~Heinke and A.~Y.~Potekhin,
%``Tests of the nuclear equation of state and superfluid and superconducting gaps using the Cassiopeia A neutron star,''
Phys. Rev. C \textbf{91} (2015) no.1, 015806
doi:10.1103/PhysRevC.91.015806 [arXiv:1412.7759 [astro-ph.HE]].
%56 citations counted in INSPIRE as of 24 Jan 2021


%\cite{Kanakis-Pegios:2020kzp}
\bibitem{Kanakis-Pegios:2020kzp}
A.~Kanakis-Pegios, P.~S.~Koliogiannis and C.~C.~Moustakidis,
%``Probing the nuclear equation of state from the existence of a $\sim 2.6~M_{\odot}$ neutron star: the GW190814 puzzle,''
[arXiv:2012.09580 [astro-ph.HE]].
%1 citations counted in INSPIRE as of 24 Jan 2021

%%%%%%%%%%%%%%%%%%%%%%%%%%%%%%%%%%%%%%%%%%%%%%%%%%%%%%%%%%%%%%%%%%%%%%%%%%%%%%%%%%%%%%%%%%%%%%%%nuclear


%\cite{Buschmann:2019pfp}
\bibitem{Buschmann:2019pfp}
M.~Buschmann, R.~T.~Co, C.~Dessert and B.~R.~Safdi,
%``X-ray Search for Axions from Nearby Isolated Neutron Stars,''
Phys. Rev. Lett. \textbf{126} (2021) no.2, 021102
doi:10.1103/PhysRevLett.126.021102 [arXiv:1910.04164 [hep-ph]].
%11 citations counted in INSPIRE as of 24 Jan 2021




%\cite{Safdi:2018oeu}
\bibitem{Safdi:2018oeu}
B.~R.~Safdi, Z.~Sun and A.~Y.~Chen,
%``Detecting Axion Dark Matter with Radio Lines from Neutron Star Populations,''
Phys. Rev. D \textbf{99} (2019) no.12, 123021
doi:10.1103/PhysRevD.99.123021 [arXiv:1811.01020 [astro-ph.CO]].
%37 citations counted in INSPIRE as of 24 Jan 2021



%\cite{Hook:2018iia}
\bibitem{Hook:2018iia}
A.~Hook, Y.~Kahn, B.~R.~Safdi and Z.~Sun,
%``Radio Signals from Axion Dark Matter Conversion in Neutron Star  Magnetospheres,''
Phys. Rev. Lett. \textbf{121} (2018) no.24, 241102
doi:10.1103/PhysRevLett.121.241102 [arXiv:1804.03145 [hep-ph]].
%49 citations counted in INSPIRE as of 24 Jan 2021


%\cite{Edwards:2020afl}
\bibitem{Edwards:2020afl}
T.~D.~P.~Edwards, B.~J.~Kavanagh, L.~Visinelli and C.~Weniger,
%``Transient Radio Signatures from Neutron Star Encounters with QCD Axion Miniclusters,''
[arXiv:2011.05378 [hep-ph]].
%2 citations counted in INSPIRE as of 24 Jan 2021




%\cite{Sedrakian:2015krq}
\bibitem{Sedrakian:2015krq}
A.~Sedrakian,
%``Axion cooling of neutron stars,''
Phys. Rev. D \textbf{93} (2016) no.6, 065044
doi:10.1103/PhysRevD.93.065044 [arXiv:1512.07828 [astro-ph.HE]].
%49 citations counted in INSPIRE as of 19 Feb 2021

\bibitem{Khadkikar:2021yrj}
  S.~Khadkikar, A.~R.~Raduta, M.~Oertel and A.~Sedrakian,
  %``Maximum mass of compact stars from gravitational wave events with finite-temperature equations of state,''
  arXiv:2102.00988 [astro-ph.HE].


%\cite{Sedrakian:2006zza}
\bibitem{Sedrakian:2006zza}
D.~M.~Sedrakian, M.~V.~Hayrapetyan and M.~K.~Shahabasyan,
%``Gravitational radiation of slowly rotating neutron stars,''
Astrophysics \textbf{49} (2006), 194-200
doi:10.1007/s10511-006-0020-4
%3 citations counted in INSPIRE as of 19 Feb 2021


%\cite{Sedrakian:2018kdm}
\bibitem{Sedrakian:2018kdm}
A.~Sedrakian,
%``Axion cooling of neutron stars. II. Beyond hadronic axions,''
Phys. Rev. D \textbf{99} (2019) no.4, 043011
doi:10.1103/PhysRevD.99.043011 [arXiv:1810.00190 [astro-ph.HE]].
%13 citations counted in INSPIRE as of 19 Feb 2021







%\cite{Bauswein:2020kor}
\bibitem{Bauswein:2020kor}
A.~Bauswein, G.~Guo, J.~H.~Lien, Y.~H.~Lin and M.~R.~Wu,
%``Compact Dark Objects in Neutron Star Mergers,''
[arXiv:2012.11908 [astro-ph.HE]].
%0 citations counted in INSPIRE as of 24 Jan 2021



%\cite{Vretinaris:2019spn}
\bibitem{Vretinaris:2019spn}
S.~Vretinaris, N.~Stergioulas and A.~Bauswein,
%``Empirical relations for gravitational-wave asteroseismology of binary neutron star mergers,''
Phys. Rev. D \textbf{101} (2020) no.8, 084039
doi:10.1103/PhysRevD.101.084039 [arXiv:1910.10856 [gr-qc]].
%7 citations counted in INSPIRE as of 24 Jan 2021



%\cite{Bauswein:2020aag}
\bibitem{Bauswein:2020aag}
A.~Bauswein, S.~Blacker, V.~Vijayan, N.~Stergioulas,
K.~Chatziioannou, J.~A.~Clark, N.~U.~F.~Bastian, D.~B.~Blaschke,
M.~Cierniak and T.~Fischer,
%``Equation of state constraints from the threshold binary mass for prompt collapse of neutron star mergers,''
Phys. Rev. Lett. \textbf{125} (2020) no.14, 141103
doi:10.1103/PhysRevLett.125.141103 [arXiv:2004.00846
[astro-ph.HE]].
%19 citations counted in INSPIRE as of 24 Jan 2021


%\cite{Bauswein:2017vtn}
\bibitem{Bauswein:2017vtn}
A.~Bauswein, O.~Just, H.~T.~Janka and N.~Stergioulas,
%``Neutron-star radius constraints from GW170817 and future detections,''
Astrophys. J. Lett. \textbf{850} (2017) no.2, L34
doi:10.3847/2041-8213/aa9994 [arXiv:1710.06843 [astro-ph.HE]].
%276 citations counted in INSPIRE as of 24 Jan 2021








%\cite{Haensel:2007yy}
\bibitem{Haensel:2007yy}
P.~Haensel, A.~Y.~Potekhin and D.~G.~Yakovlev,
%``Neutron stars 1: Equation of state and structure,''
Astrophys. Space Sci. Libr. \textbf{326} (2007), pp.1-619
doi:10.1007/978-0-387-47301-7
%192 citations counted in INSPIRE as of 19 Dec 2020




%\cite{Friedman:2013xza}
\bibitem{Friedman:2013xza}
J.~L.~Friedman and N.~Stergioulas,
%``Rotating Relativistic Stars,''
doi:10.1017/CBO9780511977596
%27 citations counted in INSPIRE as of 19 Dec 2020

%\cite{Baym:2017whm}
\bibitem{Baym:2017whm}
G.~Baym, T.~Hatsuda, T.~Kojo, P.~D.~Powell, Y.~Song and
T.~Takatsuka,
%``From hadrons to quarks in neutron stars: a review,''
Rept. Prog. Phys. \textbf{81} (2018) no.5, 056902
doi:10.1088/1361-6633/aaae14 [arXiv:1707.04966 [astro-ph.HE]].
%213 citations counted in INSPIRE as of 19 Dec 2020


%\cite{Lattimer:2004pg}
\bibitem{Lattimer:2004pg}
J.~M.~Lattimer and M.~Prakash,
%``The physics of neutron stars,''
Science \textbf{304} (2004), 536-542 doi:10.1126/science.1090720
[arXiv:astro-ph/0405262 [astro-ph]].
%814 citations counted in INSPIRE as of 19 Dec 2020


%\cite{Olmo:2019flu}
\bibitem{Olmo:2019flu}
G.~J.~Olmo, D.~Rubiera-Garcia and A.~Wojnar,
%``Stellar structure models in modified theories of gravity: Lessons and challenges,''
Phys. Rept. \textbf{876} (2020), 1-75
doi:10.1016/j.physrep.2020.07.001 [arXiv:1912.05202 [gr-qc]].
%37 citations counted in INSPIRE as of 24 Jan 2021


%\cite{Read:2008iy}
\bibitem{Read:2008iy}
J.~S.~Read, B.~D.~Lackey, B.~J.~Owen and J.~L.~Friedman,
%``Constraints on a phenomenologically parameterized neutron-star equation of state,''
Phys. Rev. D \textbf{79} (2009), 124032
%doi:10.1103/PhysRevD.79.124032 [arXiv:0812.2163 [astro-ph]].
%422 citations counted in INSPIRE as of 19 Dec 2020


%\cite{Read:2009yp}
\bibitem{Read:2009yp}
J.~S.~Read, C.~Markakis, M.~Shibata, K.~Uryu, J.~D.~E.~Creighton
and J.~L.~Friedman,
%``Measuring the neutron star equation of state with gravitational wave observations,''
Phys. Rev. D \textbf{79} (2009), 124033
%doi:10.1103/PhysRevD.79.124033 [arXiv:0901.3258 [gr-qc]].
%269 citations counted in INSPIRE as of 19 Dec 2020






\bibitem{reviews1}
 S.~Nojiri, S.~D.~Odintsov and V.~K.~Oikonomou,
  %``Modified Gravity Theories on a Nutshell: Inflation, Bounce and Late-time Evolution,''
  Phys.\ Rept.\  {\bf 692} (2017) 1
  doi:10.1016/j.physrep.2017.06.001
  [arXiv:1705.11098 [gr-qc]].
  %%CITATION = doi:10.1016/j.physrep.2017.06.001;%%
  %21 citations counted in INSPIRE as of 20 Aug 2017



\bibitem{reviews2}

S. Nojiri, S.D. Odintsov,
   %``Unified cosmic history in modified gravity: from F(R) theory to
   %Lorentz non-invariant models,''
   Phys.\ Rept.\  {\bf 505}, 59 (2011);
   %[arXiv:1011.0544 [gr-qc]].
   %%CITATION = ARXIV:1011.0544;%%



\bibitem{reviews3}
S. Nojiri, S.D. Odintsov,
  %``Introduction to modified gravity and gravitational alternative for dark
  %energy,''
  eConf {\bf C0602061}, 06 (2006)
  [Int.\ J.\ Geom.\ Meth.\ Mod.\ Phys.\  {\bf 4}, 115 (2007)].
 [arXiv:hep-th/0601213];
  %%CITATION = 00436,4,115;%%


   \bibitem{reviews4}
 S. Capozziello, M. De Laurentis,
   %``Extended Theories of Gravity,''
   Phys.\ Rept.\  {\bf 509}, 167 (2011)
   [arXiv:1108.6266 [gr-qc]].
   %%CITATION = ARXIV:1108.6266;%%

   \bibitem{book}
  S.~Capozziello, V. Faraoni
 \textit{Beyond Einstein Gravity : A Survey of Gravitational Theories for Cosmology and Astrophysics},
  Fundam.\ Theor.\ Phys.\  {\bf 170}, Springer (2011), Dordrecht.\\
  doi:10.1007/978-94-007-0165-6
  %%CITATION = doi:10.1007/978-94-007-0165-6;%%
  %64 citations counted in INSPIRE as of 16 Sep 2017



\bibitem{reviews5}

A.~de la Cruz-Dombriz and D.~Saez-Gomez,
  %``Black holes, cosmological solutions, future singularities, and their thermodynamical properties in modified gravity theories,''
  Entropy {\bf 14} (2012) 1717
  doi:10.3390/e14091717
  [arXiv:1207.2663 [gr-qc]].
  %%CITATION = doi:10.3390/e14091717;%%
  %125 citations counted in INSPIRE as of 15 Nov 2017

\bibitem{reviews6}

G.~J.~Olmo,
  %``Palatini Approach to Modified Gravity: f(R) Theories and Beyond,''
  Int.\ J.\ Mod.\ Phys.\ D {\bf 20} (2011) 413
  doi:10.1142/S0218271811018925
  [arXiv:1101.3864 [gr-qc]].
  %%CITATION = doi:10.1142/S0218271811018925;%%
  %204 citations counted in INSPIRE as of 04 Feb 2018




\bibitem{dimo} K. Dimopoulos, \textit{Introduction to Cosmic Inflation and Dark Energy}, (2021) CRC Press




%\cite{Astashenok:2014pua}
\bibitem{Astashenok:2014pua}
A.~V.~Astashenok, S.~Capozziello and S.~D.~Odintsov,
%``Maximal neutron star mass and the resolution of the hyperon puzzle in modified gravity,''
Phys. Rev. D \textbf{89} (2014) no.10, 103509
doi:10.1103/PhysRevD.89.103509 [arXiv:1401.4546 [gr-qc]].
%117 citations counted in INSPIRE as of 19 Dec 2020

\bibitem{Birrell}
N. D. Birrell and  P. C. W. Davies, \textit{Quantum fields in Curved Space}, Cambridge Univ. Press (1982) Cambridge,\\
doi:10.1017/CBO9780511622632

\bibitem{Mauro}
%\cite{Capozziello:2007ec}
%\bibitem{Capozziello:2007ec}
  S.~Capozziello and M.~Francaviglia,
  %``Extended Theories of Gravity and their Cosmological and Astrophysical Applications,''
  Gen.\ Rel.\ Grav.\  {\bf 40} (2008) 357
  doi:10.1007/s10714-007-0551-y
  [arXiv:0706.1146 [astro-ph]].
  %%CITATION = doi:10.1007/s10714-007-0551-y;%%
  %671 citations counted in INSPIRE as of 27 Feb 2021

\bibitem{Curv}
%\cite{Capozziello:2006uv}
%\bibitem{Capozziello:2006uv}
  S.~Capozziello, V.~F.~Cardone and A.~Troisi,
  %``Dark energy and dark matter as curvature effects,''
  JCAP {\bf 0608} (2006) 001,
  doi:10.1088/1475-7516/2006/08/001
  [astro-ph/0602349].
  %%CITATION = doi:10.1088/1475-7516/2006/08/001;%%
  %242 citations counted in INSPIRE as of 27 Feb 2021


%\cite{Douchin:2001sv}
\bibitem{Douchin:2001sv}
F.~Douchin and P.~Haensel,
%``A unified equation of state of dense matter and neutron star structure,''
Astron. Astrophys. \textbf{380} (2001), 151
doi:10.1051/0004-6361:20011402 [arXiv:astro-ph/0111092
[astro-ph]].
%625 citations counted in INSPIRE as of 19 Dec 2020









%\cite{Abbott:2020khf}
\bibitem{Abbott:2020khf}
R.~Abbott \textit{et al.} [LIGO Scientific and Virgo],
%``GW190814: Gravitational Waves from the Coalescence of a 23 Solar Mass Black Hole with a 2.6 Solar Mass Compact Object,''
Astrophys. J. Lett. \textbf{896} (2020) no.2, L44
doi:10.3847/2041-8213/ab960f [arXiv:2006.12611 [astro-ph.HE]].
%234 citations counted in INSPIRE as of 19 Dec 2020



%\cite{Roupas:2020nua}
\bibitem{Roupas:2020nua}
Z.~Roupas, G.~Panotopoulos and I.~Lopes,
%``QCD color superconductivity in compact stars: color-flavor locked quark star candidate for the gravitational-wave signal GW190814,''
[arXiv:2010.11020 [astro-ph.HE]].
%6 citations counted in INSPIRE as of 09 Mar 2021




%\cite{Biswas:2020xna}
\bibitem{Biswas:2020xna}
B.~Biswas, R.~Nandi, P.~Char, S.~Bose and N.~Stergioulas,
%``GW190814: On the properties of the secondary component of the binary,''
[arXiv:2010.02090 [astro-ph.HE]].
%6 citations counted in INSPIRE as of 19 Dec 2020


  %\cite{Bombaci:2020vgw}
\bibitem{Bombaci}
  I.~Bombaci, A.~Drago, D.~Logoteta, G.~Pagliara and I.~Vidana,
  %``Was GW190814 a black hole -- strange quark star system?,''
  arXiv:2010.01509 [nucl-th].
  %%CITATION = ARXIV:2010.01509;%%
  %11 citations counted in INSPIRE as of 27 Feb 2021


%\cite{Rhoades:1974fn}
\bibitem{Rhoades:1974fn}
C.~E.~Rhoades, Jr. and R.~Ruffini,
%``Maximum mass of a neutron star,''
Phys. Rev. Lett. \textbf{32} (1974), 324-327
doi:10.1103/PhysRevLett.32.324
%376 citations counted in INSPIRE as of 13 Dec 2020



%\cite{Kalogera:1996ci}
\bibitem{Kalogera:1996ci}
V.~Kalogera and G.~Baym,
%``The maximum mass of a neutron star,''
Astrophys. J. Lett. \textbf{470} (1996), L61-L64
doi:10.1086/310296 [arXiv:astro-ph/9608059 [astro-ph]].
%194 citations counted in INSPIRE as of 19 Dec 2020




%\cite{Bailyn:1997xt}
\bibitem{Bailyn:1997xt}
C.~D.~Bailyn, R.~K.~Jain, P.~Coppi and J.~A.~Orosz,
%``The Mass distribution of stellar black holes,''
Astrophys. J. \textbf{499} (1998), 367 doi:10.1086/305614
[arXiv:astro-ph/9708032 [astro-ph]].
%146 citations counted in INSPIRE as of 19 Dec 2020


%\cite{Fasano:2020eum}
\bibitem{Fasano:2020eum}
M.~Fasano, K.~W.~K.~Wong, A.~Maselli, E.~Berti, V.~Ferrari and
B.~S.~Sathyaprakash,
%``Distinguishing double neutron star from neutron star-black hole binary populations with gravitational wave observations,''
Phys. Rev. D \textbf{102} (2020) no.2, 023025
doi:10.1103/PhysRevD.102.023025 [arXiv:2005.01726 [astro-ph.HE]].
%8 citations counted in INSPIRE as of 19 Dec 2020





%\cite{Astashenok:2020qds}
\bibitem{Astashenok:2020qds}
A.~V.~Astashenok, S.~Capozziello, S.~D.~Odintsov and
V.~K.~Oikonomou,
%``Extended Gravity Description for the GW190814 Supermassive Neutron Star,''
Phys. Lett. B \textbf{811} (2020), 135910
doi:10.1016/j.physletb.2020.135910 [arXiv:2008.10884 [gr-qc]].
%6 citations counted in INSPIRE as of 19 Dec 2020





%\cite{Capozziello:2015yza}
\bibitem{Capozziello:2015yza}
S.~Capozziello, M.~De Laurentis, R.~Farinelli and S.~D.~Odintsov,
%``Mass-radius relation for neutron stars in f(R) gravity,''
Phys. Rev. D \textbf{93} (2016) no.2, 023501
doi:10.1103/PhysRevD.93.023501 [arXiv:1509.04163 [gr-qc]].
%116 citations counted in INSPIRE as of 19 Dec 2020



%\cite{Astashenok:2014nua}
\bibitem{Astashenok:2014nua}
A.~V.~Astashenok, S.~Capozziello and S.~D.~Odintsov,
%``Extreme neutron stars from Extended Theories of Gravity,''
JCAP \textbf{01} (2015), 001 doi:10.1088/1475-7516/2015/01/001
[arXiv:1408.3856 [gr-qc]].
%109 citations counted in INSPIRE as of 19 Dec 2020



%\cite{Astashenok:2013vza}
\bibitem{Astashenok:2013vza}
A.~V.~Astashenok, S.~Capozziello and S.~D.~Odintsov,
%``Further stable neutron star models from f(R) gravity,''
JCAP \textbf{12} (2013), 040 doi:10.1088/1475-7516/2013/12/040
[arXiv:1309.1978 [gr-qc]].
%142 citations counted in INSPIRE as of 19 Dec 2020


%\cite{DeLaurentis:2018odx}
\bibitem{Laurentis}
  M.~De Laurentis,
  %``Noether's stars in $f(\cal {R})$ gravity,''
  Phys.\ Lett.\ B {\bf 780} (2018) 205,
  doi:10.1016/j.physletb.2018.03.001
  [arXiv:1802.09073 [gr-qc]].
  %%CITATION = doi:10.1016/j.physletb.2018.03.001;%%
  %4 citations counted in INSPIRE as of 27 Feb 2021

%\cite{Pani:2014jra}
\bibitem{Pani:2014jra}
P.~Pani and E.~Berti,
%``Slowly rotating neutron stars in scalar-tensor theories,''
Phys. Rev. D \textbf{90} (2014) no.2, 024025
doi:10.1103/PhysRevD.90.024025 [arXiv:1405.4547 [gr-qc]].
%86 citations counted in INSPIRE as of 19 Dec 2020




%\cite{Ramazanoglu:2016kul}
\bibitem{Ramazanoglu:2016kul}
F.~M.~Ramazano\u{g}lu and F.~Pretorius,
%``Spontaneous Scalarization with Massive Fields,''
Phys. Rev. D \textbf{93} (2016) no.6, 064005
doi:10.1103/PhysRevD.93.064005 [arXiv:1601.07475 [gr-qc]].
%83 citations counted in INSPIRE as of 19 Dec 2020


%\cite{Yazadjiev:2014cza}
\bibitem{Yazadjiev:2014cza}
S.~S.~Yazadjiev, D.~D.~Doneva, K.~D.~Kokkotas and K.~V.~Staykov,
%``Non-perturbative and self-consistent models of neutron stars in R-squared gravity,''
JCAP \textbf{06} (2014), 003 doi:10.1088/1475-7516/2014/06/003
[arXiv:1402.4469 [gr-qc]].
%82 citations counted in INSPIRE as of 19 Dec 2020




%\cite{Berti:2020kgk}
\bibitem{Berti:2020kgk}
E.~Berti, L.~G.~Collodel, B.~Kleihaus and J.~Kunz,
%``Spin-induced black-hole scalarization in Einstein-scalar-Gauss-Bonnet theory,''
[arXiv:2009.03905 [gr-qc]].
%9 citations counted in INSPIRE as of 19 Dec 2020





%\cite{Silva:2017uqg}
\bibitem{Silva:2017uqg}
H.~O.~Silva, J.~Sakstein, L.~Gualtieri, T.~P.~Sotiriou and
E.~Berti,
%``Spontaneous scalarization of black holes and compact stars from a Gauss-Bonnet coupling,''
Phys. Rev. Lett. \textbf{120} (2018) no.13, 131104
doi:10.1103/PhysRevLett.120.131104 [arXiv:1711.02080 [gr-qc]].
%190 citations counted in INSPIRE as of 19 Dec 2020


%\cite{Pani:2011xm}
\bibitem{Pani:2011xm}
P.~Pani, E.~Berti, V.~Cardoso and J.~Read,
%``Compact stars in alternative theories of gravity. Einstein-Dilaton-Gauss-Bonnet gravity,''
Phys. Rev. D \textbf{84} (2011), 104035
doi:10.1103/PhysRevD.84.104035 [arXiv:1109.0928 [gr-qc]].
%102 citations counted in INSPIRE as of 19 Dec 2020





%\cite{Astashenok:2020cfv}
\bibitem{Astashenok:2020cfv}
A.~V.~Astashenok and S.~D.~Odintsov,
%``Supermassive Neutron Stars in Axion $F(R)$ Gravity,''
Mon. Not. Roy. Astron. Soc. \textbf{493} (2020) no.1, 78-86
doi:10.1093/mnras/staa214 [arXiv:2001.08504 [gr-qc]].
%4 citations counted in INSPIRE as of 19 Dec 2020




%\cite{Day:2019bbh}
\bibitem{Day:2019bbh}
F.~V.~Day and J.~I.~McDonald,
%``Axion superradiance in rotating neutron stars,''
JCAP \textbf{10} (2019), 051 doi:10.1088/1475-7516/2019/10/051
[arXiv:1904.08341 [hep-ph]].
%18 citations counted in INSPIRE as of 19 Dec 2020






%\cite{Brito:2017zvb}
\bibitem{Brito:2017zvb}
R.~Brito, S.~Ghosh, E.~Barausse, E.~Berti, V.~Cardoso, I.~Dvorkin,
A.~Klein and P.~Pani,
%``Gravitational wave searches for ultralight bosons with LIGO and LISA,''
Phys. Rev. D \textbf{96} (2017) no.6, 064050
doi:10.1103/PhysRevD.96.064050 [arXiv:1706.06311 [gr-qc]].
%123 citations counted in INSPIRE as of 19 Dec 2020







%\cite{Sotani:2017pfj}
\bibitem{Sotani:2017pfj}
H.~Sotani and K.~D.~Kokkotas,
%``Maximum mass limit of neutron stars in scalar-tensor gravity,''
Phys. Rev. D \textbf{95} (2017) no.4, 044032
doi:10.1103/PhysRevD.95.044032 [arXiv:1702.00874 [gr-qc]].
%20 citations counted in INSPIRE as of 19 Dec 2020




\bibitem{weinberg}S.Weinberg "{\it Gravitation and Cosmology}", John Wiley \&
Sons, Inc., New York, (1972).

\bibitem{capquark}A. V. Astashenok, S. Capozziello, S. D. Odintsov, {\it Physics Letters B} {\bf 742}, 160 (2015).

\bibitem{rezzollazan}L. Rezzolla, O. Zanotti, "{\it Relativistic hydrodinamics}", Oxford University Press, Oxford UK (2013).

\bibitem{landaufluid}L.D. Landau, E.M. Lifshitz, "{\it Fluid Mechanics}" Course of theoretical physics vol. 6, Butterworth-Heinemann. Oxford (1987).

%%%%%%%%%%%%%%%%%%%%%%%%%%%%%%%%%%%%%%%%%%%%%%%%%%%%%%%%%%%%%%%%%%%%%%%%%%%%%%%%%%%%%%%%%%%%%%%%%%%%%%%%OK
 %\cite{Friedman:1990qz}
\bibitem{Friedman:1990qz}
J.~L.~Friedman and R.~R.~Caldwell,
%``Evidence against a strange ground state for baryons,''
Phys. Lett. B \textbf{264} (1991), 143-148
doi:10.1016/0370-2693(91)90718-6
%82 citations counted in INSPIRE as of 23 Feb 2021





%\cite{Foucart:2018rjc}
\bibitem{Foucart:2018rjc}
F.~Foucart, T.~Hinderer and S.~Nissanke,
%``Remnant baryon mass in neutron star-black hole mergers: Predictions for binary neutron star mimickers and rapidly spinning black holes,''
Phys. Rev. D \textbf{98} (2018) no.8, 081501
doi:10.1103/PhysRevD.98.081501 [arXiv:1807.00011 [astro-ph.HE]].
%75 citations counted in INSPIRE as of 19 Dec 2020





%\cite{Zappa:2019ntl}
\bibitem{Zappa:2019ntl}
F.~Zappa, S.~Bernuzzi, F.~Pannarale, M.~Mapelli and N.~Giacobbo,
%``Black-Hole Remnants from Black-Hole\textendash{}Neutron-Star Mergers,''
Phys. Rev. Lett. \textbf{123} (2019) no.4, 041102
doi:10.1103/PhysRevLett.123.041102 [arXiv:1903.11622 [gr-qc]].
%14 citations counted in INSPIRE as of 19 Dec 2020



%\cite{Foucart:2019bxj}
\bibitem{Foucart:2019bxj}
F.~Foucart, M.~D.~Duez, L.~E.~Kidder, S.~Nissanke, H.~P.~Pfeiffer
and M.~A.~Scheel,
%``Numerical simulations of neutron star-black hole binaries in the near-equal-mass regime,''
Phys. Rev. D \textbf{99} (2019) no.10, 103025
doi:10.1103/PhysRevD.99.103025 [arXiv:1903.09166 [astro-ph.HE]].
%31 citations counted in INSPIRE as of 19 Dec 2020









\end{thebibliography}
\end{document}